# Carbon deposition on hematite ($\alpha$-Fe$_2$O$_3$) nanocubes by an annealing in the air: morphology study with grazing incidence small angle X-ray scattering (GISAXS)

**Chang-Yong Kim [1],***

[1] Canadian Light Source, 44 Innovation boulevard, Saskatoon, SK S7N 2V3 Canada
* Correspondence: Chang-Yong.Kim@lightsource.ca

**Abstract**

GISAXS has been used to study morphology change of α-Fe$_2$O$_3$ nanocubes after annealing processes. A submonolayer of the nanocubes was deposited on a Si(100) substrate. While an annealing at 400 °C in a vacuum does not change a GISAXS pattern from as-prepared nanocubes submonolayer, subsequent annealing in the air at the same temperature altered the GISAXS pattern significantly. SEM image showed that the air-annealed nanocubes were coated with thin layers which were identified as amorphous carbon layers based on Raman measurements. GISAXS simulations from morphologies of nanocube with 38 nm side-length and core-shell (nanocube-core and 7 nm thick carbon-shell) reproduced measured patterns from the vacuum annealed and the air-annealed nanocubes, respectively. The current study provides new approach for in-situ characterization of carbon deposition on a uniform shape nanoparticle through monitoring of deposited carbon thickness.

## 1. Introduction

Hematite ($\alpha$-Fe$_2$O$_3$) is strong candidate for photocatalytic applications because its band gap (2.1 eV) can give the maximum theoretical solar-to-hydrogen efficiency of ~15% and is corrosion-resistive to harsh oxidation condition [1,2]. For photocatalytic application a good crystalline nanoparticle with size of a few tens nanometers is preferred because light absorption and formation of depletion layer require a thick hematite material [3]. For practical applications an as-prepared catalyst in a nanoparticle form undergoes thermal treatments for crystallization, structural transform and activation. Often the thermal treatment introduces morphology changes or agglomeration of nanoparticles and subsequently affects performance the catalysts. However, preserve nanostructure could be essential to maintain an activity [4].

Numerous monodispersive uniform-shape hematite nanoparticles have been synthesized [5-14]. Monodispersive and uniform-shape nanoparticles provide unique opportunity to study morphology change of the nanoparticles during a chemical reaction. Subtle structural changes of these nanoparticles could be detected relatively easily because shape and size of nanoparticles prior to a reaction are well defined. Especially a pseudo-cubic hematite nanoparticle (hereafter called as nanocube) is enclosed by six crystallographically equivalent $\{01\bar{1}2\}$ faces [15].

A performance of hematite in practical photocatalytic application is far behind from the theoretical prediction [2]. Numerous limiting factors have been identified, for example poor conductivity [16], short hole diffusion length [17], and existence of surface trap state [18]. Especially a passivation of the surface trap state through a surface layer coating enhanced photocatalytic performance [19,20]. There was report that carbon coating of hematite nanoparticle also resulted about fourfold increase of photocurrent compare to that of bare hematite without carbon coating [21]. Similar carbon coatings were achieved by pyrolysis of dopamine [22], hydrothermal synthesis with dimethyl diallyl ammonium chloride additive [23], carbonization of the n-butane (fire treatment) [24]. However, current study shows that it is possible to coat carbon on the hematite nanocube by annealing in the air without using any agent at all.

It would be very useful to correlate catalytic performance to thickness of coated carbon on a specific crystallographic facet of hematite nanoparticle. Grazing incidence X-ray small angle scattering (GISAXS) measurement has been a powerful tool to characterize morphology and spatial distribution of nanoparticles deposited on a substrate [25,26]. If the effect from the spatial distribution is negligible, GISAXS intensities are proportional to Fourier transform of density distribution of the deposited nanoparticles. By using a submonolayer film of monodispersive hematite nanocubes we applied the GISAXS to characterize the carbon deposition on the $\{01\bar{1}2\}$ faces of the hematite nanocubes by air annealing and its effect to particle morphology.

## 2. Results and Discussion

In a submonolayer of the nanocubes formed on a substrate, the nanocube would adsorb with its faces, not edge nor vertex, contacting with the substrate. Consequently, the top surfaces of all nanocubes are parallel to the substrate and their in-plane orientations are random [10,15]. GISAXS measurements were taken from submonolayer films of the as-prepared, annealed in vacuum and subsequently annealed in the air. GISAXS patterns from the as-prepared and the vacuum annealed films were very similar and GISAXS pattern from vacuum annealed film is shown in Figure 1.

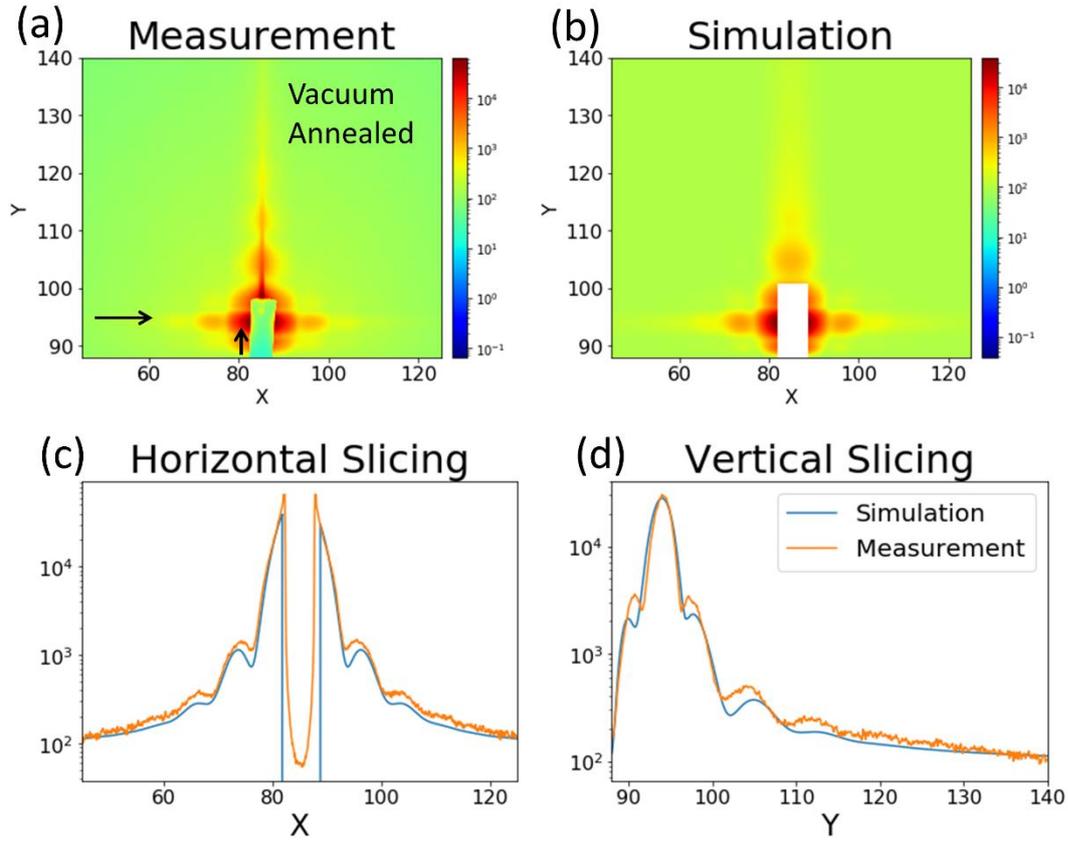

**Figure 1.** (a) GISAXS patterns taken from nanocube submolayer annealed in vacuum, (b) simulated pattern, comparison of measurements and simulations along (c) horizontal and (d) vertical slices. Arrows in (a) indicate positions where slices of (c) and (d) were taken.

Intensities $I(\vec{q})$ in GISAXS image can be represented by following equations.

$$I(\vec{q}) = \langle |F|^2 \rangle S(q_{\parallel})$$

$$F(\vec{q}) = \int \rho(\vec{r}) exp(i\vec{q} \cdot \vec{r}) d^3 r$$

where $\vec{q}$, $S(q_{\parallel})$, and $F(\vec{q})$ are momentum transfer, structure factor, and form factor of individual particle, respectively [25,26]. In the prepared submonolayer film, distances between nanocubes are large enough so that the structural factor, $S(q_{\parallel})$, describing interference originated from arrangement of nanocubes can be neglected. Then, GISAXS intensities are proportional to the form factor $F(\vec{q})$ of a nanocube, which is Fourier transform of density distribution of the nanocube. If the density inside nanocube is uniform, size of nanocube determines the GISAXS pattern. The GISAXS intensities were simulated with BornAgain [27]. As shown in the figure 1 the observed pattern matches well with GISAXS simulation based on the cube shape with average size of 38 nm and size distribution of ±3 nm.

The GISAXS pattern changed dramatically after subsequent annealing in air (Figure 2). It is worth to note that the nanocubes were annealed in vacuum prior to the air annealing. Thus, any chemicals involved in the nanocube synthesis were removed before the air annealing process. The change in GISAXS pattern indicates modification of shape and/or density distribution.

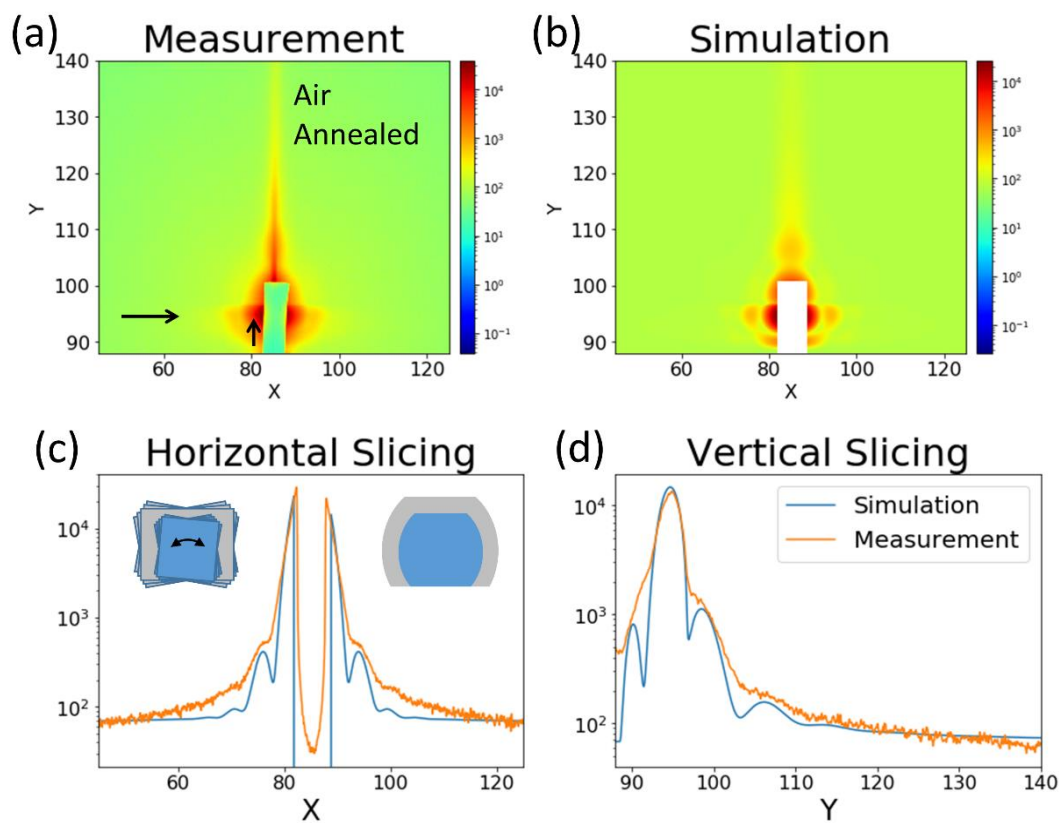

**Figure 2.** (a) GISAXS patterns taken from nanocube submolayer annealed in the air at 400 °C, (b) simulated pattern, comparison of measurement and simulation along (c) horizontal and (d) vertical

slices. Arrows in (a) indicate positions where slices of (c) and (d) were taken. Insets in (c) represents nanocube core-shells with tilting (left) and truncated sphere core-shell model used for the simulation (right).

Numerous morphologies of nanoparticle enclosed by (104), (110) and (001) crystallographic faces in addition to the (012) can be formed under various water and oxygen partial pressure during nanoparticle formation [28]. Although there is possibility of morphology change from nanocube to other shape, annealing temperature 400 °C seems too low to induce morphology change. To check actual morphologies of nanocubes after vacuum and air annealing SEM was used as shown in Figure 3. Vacuum annealed nanocubes show clear cubic morphology. The SEM image showed that the air annealed nanocubes were encapsulated by a thin layer of a low-Z material.

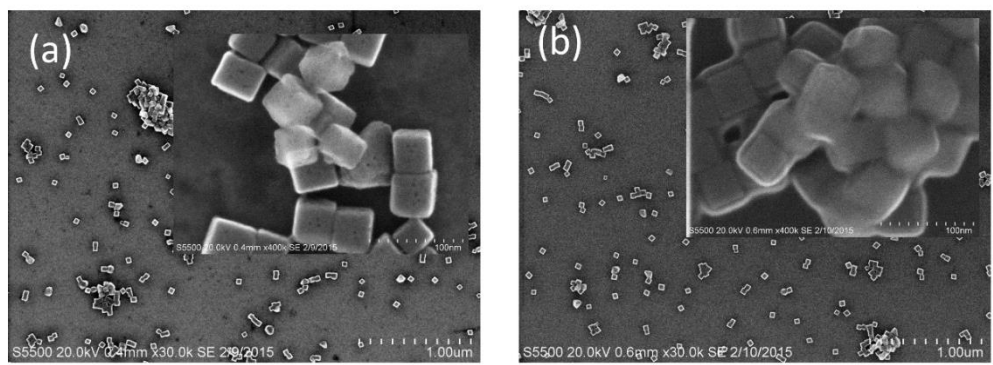

**Figure 3.** (a) SEM images from vacuum annealed and (b) air-annealed nanocubes. Insets show images with bigger magnification (10X).

The capping layer was identified as an amorphous carbon through Raman measurement as shown in Figure 4. It is well known through $CO_2$ hydrogenation study that carbon deposition on hematite catalyst is a major reaction at reaction temperature of 400 °C [29]. The sharp strong peak at 520.7 $cm^{-1}$ and a broad peak in the range of 930 – 1030 $cm^{-1}$ are from the silicon substrate [30]. The peak around 2330 cm$^{-1}$ is from nitrogen molecules in the air [31]. Raman peaks from hematite agree well with literature [32,33]. The weak and broad peaks around 2610 and 2900 cm$^{-1}$ are from hematite [33]. The broad peak around 1550 cm$^{-1}$ is typical G-peak of disordered carbon [34] and D-peak (around 1350 cm$^{-1}$) might be buried under strong hematite peak at 1310 cm$^{-1}$. As disorder increases the carbon D-peak intensity increases [34] and higher order peaks of D-peak get weaker too. Hence, the absence (or very weak) of the second order bands around 2500 - 3500 cm$^{-1}$ indicates that the capping layer is in an amorphous carbon form [35].

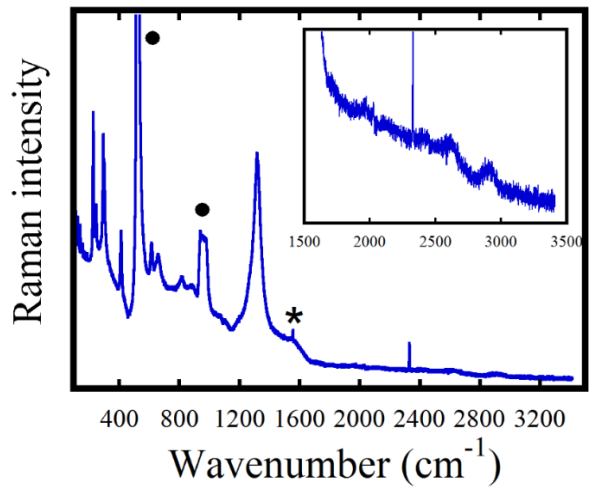

**Figure 4.** Raman spectrum measured from the air annealed nanocube submonolayer. The sharp strong peak at 520.7 cm$^{-1}$ and broad peak at 930 – 1030 cm$^{-1}$ marked by filled circle are from silicon substrate. The * marks carbon G band.

Based on the Raman measurement, the GISAXS pattern from the air annealed nanocube was simulated with core-shell particle shape comprising hematite nanocube core and carbon shell. Carbon shell was added as a box shape over the hematite nanocube core. It was reported that reduction of hematite coated with carbon film started above 650 °C [36]. It is not expected to have density change in hematite core through reduction by deposited carbon with annealing at 400 °C.

Hence, the size of the nanocube core was kept the same as that of the vacuum annealed nanocube during the air annealed GISAXS simulation. The nanocube core might firmly contact silicon substrate through vacuum annealing but a thin carbon layer can penetrate to the interface between nanocube and substrate. The carbon layer at the interface might not have uniform thickness. For example, one bottom edge of nanocube has thicker carbon layer than the other edges. This non-even thickness will cause a tilting of core-shell as shown at left inset in Figure 2(c). The tilting of core-shell structure cannot be handled properly with BornAgain program. Hence, the tilted and azimuthally averaged core-shell structure was approximated with a truncated sphere shape shown at the right inset in Figure 2(c). Although the approximation is crude, the existence of round edge in the truncated sphere reproduced main features of GISAXS pattern from the annealed naocubes. Simulation of GISAXS without the tilting resulted in a poor match with the measured pattern. Average thickness of carbon shell of 7 nm with size distribution of ±3 nm reproduced GISAXS pattern relatively well.

As the air annealing was performed after vacuum annealing, carbon deposition through the air annealing indicates that clean hematite nanocube surfaces interact with $CO_2$ in the air and extract carbon as amorphous form. It might be related to Fe vacancy on the nanocube surface and atomically rough surface after vacuum annealing [15]. It would be interesting to extend current approach to $CO_2$ hydrogenation [37]. Since the average thickness of deposited carbon can be extracted from GISAXS pattern, it will be possible to perform in-situ measurement to monitor carbon thickness as function of annealing time/temperature. Post-deposition of carbon over mono-dispersive uniform shape hematite nanoparticles through air annealing could provide an opportunity to study facet-dependent photo-catalytic reactions. Morphology change by reduction of hematite to magnetite through annealing at a high temperature with deposited carbon could be studied with anomalous GISAXS method [38].

In summary, GISAXS, SEM, and Raman spectroscopy have been used to characterize morphology changes of hematite nanocube resulted from vacuum and air annealing. Nanocube morphology was intact after annealing at 400 °C in vacuum. Air-annealed nanocubes were coated with thin layers which were identified as amorphous carbon. Current study can be extended to in-situ characterization of carbon deposition as function of annealing time or temperature through GISAXS measurement.

## 3. Materials and Methods

**Synthesis α-$Fe_2O_3$ nanocube**

Hematite nanocubes were prepared according to the method depicted in the Ref. [6]. Briefly, 0.721 g of $Fe(NO_3)_3 \cdot 9H_2O$ and 1.071 g of poly(Nvinyl-2-pyrrolidone (PVP, Mw = 29,000) were dissolved in 64.3 ml of N,N-dimethyformaide (DMF), then placed in a Teflon-lined stainless stain autoclave of volume 125 ml. The sealed vessel was put into an oven and heated at 180 °C for 30 hours, and then cooled to room temperature naturally. The red precipitates were collected by centrifuging, washing with deionized water four times and drying in air at room temperature. α-$Fe_2O_3$ colloids were prepared by suspending this α-$Fe_2O_3$ powder in 50 ml anhydrous toluene and ultrasonicating under nitrogen for 1 hour.

**Fabrication of submonolayer of α-$Fe_2O_3$ nanocube**

Submonolayer of α-$Fe_2O_3$ nanocube was fabricated following the procedure in Ref. [10] after a slight modification. Silicon substrates were cleanned by annealing at 700 °C for 3 hours under $O_2$ flow and cooled down to room temperature while keeping the $O_2$ flow. The substrates were then immersed in deionized water at 90 $^0$C for 1 hour, rinsed with dilute HCl and $H_2O$, and then blown-dry with $N_2$. The substrates were submerged into a solution of 4 ml hexamethylene diisocyanate (HDI) in 40 ml anhydrous toluene at 60 °C under nitrogen for 24 hours and rinsed with anhydrous toluene to remove unbound HDI from surface. Finally the substrates were immerged into the colloids of α-Fe$_2$O$_3$ and heated at 60 °C for 24 hours. These substrates were rinsed with methanol and dried under vacuum at 60°C for 2 hours.

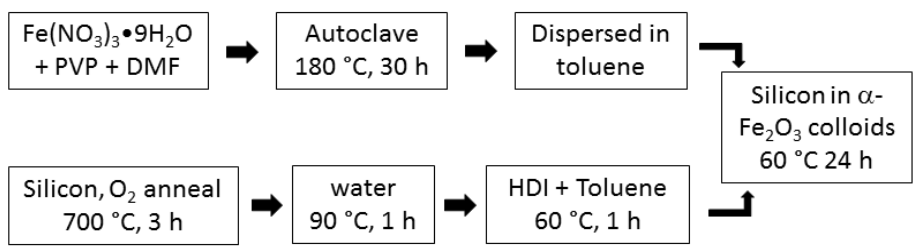

**Figure 5.** Flow scheme of α-Fe$_2$O$_3$ nanocube submonolayer sample preparation

**GISAXS measurements**

The GISAXS measurements (setup shown in the Figure 6) were performed at Hard X-ray MicroAnalysis (HXMA) beamline in Canadian Light Source. The substrate was mounted inside a small UHV chamber equipped with graphite heater embedded in pyritic boron nitride. Inside of the chamber was pumped to 1x10$^{-6}$ torr during a vacuum annealing. GISAXS measurements were performed with 8979 eV incident X-ray energy and CCD X-ray detector (Rayonix SX165) was placed 2.35 meters downstream from the sample. Between the sample and CCD detector a vacuum tube connected to the small UHV chamber was placed to minimize air scattering. The vacuum tube was directly connected to the sample chamber through a gate valve with Kapton window. The gate valve was closed during the annealing processes. For an air annealing treatment, nanocubes were vacuum-annealed at 400 °C (1x10$^{-5}$ Torr) and subsequently the air was leaked into the chamber while maintaining the substrate temperature at 400 °C. The GISAXS patterns were collected with sample at the annealing temperature.

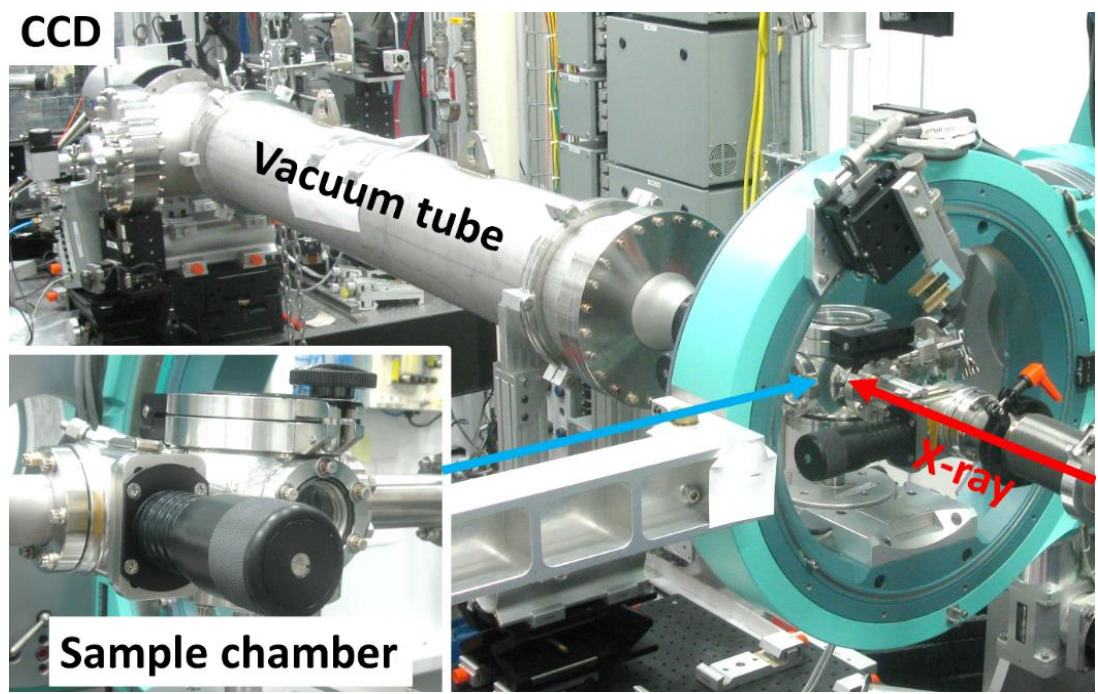

**Figure 6.** Grazing incident small angle X-ray scattering (GISAXS) setup

**Raman measurement**

Raman spectra were measured with Andor SR500i spectrograph equipped with Newton CCD camera and 532 nm laser was used as excitation source.

**Funding:** This research received no external funding.

**Acknowledgments:** The research described in this paper was performed at the Canadian Light Source, a national research facility of the University of Saskatchewan, which is supported by the Canada Foundation for Innovation (CFI), the Natural Sciences and Engineering Research Council (NSERC), the National Research Council (NRC), the Canadian Institutes of Health Research (CIHR), the Government of Saskatchewan, and the University of Saskatchewan.

**Conflicts of Interest:** The authors declare no conflict of interest.